\documentclass[twocolumn,aps,showpacs]{revtex4}
\usepackage{graphicx,amsmath,amssymb,amsfonts}
\usepackage{fmtcount}
\usepackage[utf8]{inputenc}
\usepackage[T1]{fontenc}
\begin{document}
\title{Roles of positively charged dust, ion fluid temperature, and nonthermal electrons in the formation of modified-ion-acoustic solitary and shock waves}
\author{R. K. Shikha\footnote{Corresponding author: shikha152phy@gmail.com}, M. M. Orani, and A. A. Mamun}
\affiliation{Department of Physics \& Wazed Mia Science Research Centre,
Jahangirnagar University, Savar, Dhaka-1342, Bangladesh}
\begin{abstract}
The dusty plasma system (containing nonthermally distributed inertialess electron species, warm inertial ion species, and positively charged stationary dust species) is considered. The basic features of  subsonic and supersonic modified-ion-acoustic solitary and shock waves formed in such a dusty plasma system have been investigated  by the reductive perturbation method. It has been shown that positively charged dust species plays a new role in favor of the formation of subsonic solitary and shock waves. On the other hand, the ion fluid temperature (represented by the parameter $\sigma$) and the electron nonthermal parameter (represented by $\alpha$) play new significant roles in against the formation of subsonic solitary and shock waves, and give rise to the formation of the supersonic solitary and shock waves after their ($\sigma$'s and $\alpha$'s) certain values.  It is also shown that after a certain value of the nonthermal parameter $\alpha$, the subsonic as well as supersonic  solitary and shock waves are formed with negative potential. The important applications of the results of this theoretical investigation in space and laboratory dusty plasma systems are pinpointed.
\end{abstract}
\maketitle
\section{Introduction}
There has been a great deal of renewed interest in understanding the role of positively charged dust (PCD) species   in modifying the existing features as well as in introducing new features of  linear and nonlinear propagation of the  modified-ion-acoustic (MIA) waves propagating in space  \cite{Havnes96,Gelinas98,Mendis04,Dovner94,Cairns95,Horanyi96,Mamun04,Tsintikidis96,Horanyi93,Markus99-NLCs} and laboratory  \cite{Khrapak01,Fortov03,Davletov18} dusty plasma systems, where the PCD species coexists with the electron-ion plasmas.  The dust species in such systems are positively charged  \cite{Fortov98}  due to (i) secondary emission of electrons from the dust grain surface by the impact of high energetic plasma particles like electrons or ions \cite{Chow93}, (ii) thermionic emission of electrons from the dust grain surface by the intense radiative or thermal heating \cite{Rosenberg95}, (ii) photo-emission of electrons from the dust grain surface by the interaction of high energy photons with the dust grain surface\cite{Rosenberg96}, etc.

The dispersion relation for the MIA waves in an electron-ion-PCD plasma system (containing inertialess Maxwellian electron species, inertial cold ion species, and stationary PCD species) is \cite{Mamun20,Mamun21}
\begin{eqnarray}
\frac{\omega}{kC_i}=\frac{1}{\sqrt{1+\mu+k^2\lambda_D^2}},
\label{MIA-dispersion1}
\end{eqnarray}
where $\omega$ ($k$) is the MIA  wave angular frequency (propagation constant); $C_i=(z_ik_BT_e/m_i)^{1/2}$ is  the MIA speed in which $k_B$ is the Boltzmann constant, $T_e$ is the electron temperature, and $m_i$ is the ion mass; $\lambda_D=(k_BT_e/4\pi z_in_{i0}e^2)^{1/2}$ is  the MIA wave-length scale in which $n_{i0}$ ($z_i$) is the number density (charge state) of the ion species at equilibrium, and $e$ is the magnitude of an electronic charge; $\mu=z_dn_{d0}/z_in_{i0}$ with $n_{d0}$ ($z_d$) being the number density (charge state) of the PCD species at equilibrium  for which $n_{e0}=z_dn_{d0}+z_in_{i0}$.  This means that $\mu=0$ corresponds to the electron-ion plasma, and $\mu\rightarrow\infty$ corresponds to electron-dust plasma \cite{Mamun04,Khrapak01,Fortov03,Davletov18}. Thus, $0<\mu<\infty$ is valid for the electron-ion-PCD plasmas.  The dispersion relation
(\ref{MIA-dispersion1}) for the long-wavelength limit ($k\lambda_D\ll 1$) becomes
\begin{eqnarray}
\frac{\omega}{kC_i}\simeq\sqrt{\frac{1}{1+\mu}}.
\label{MIA-dispersion2}
\end{eqnarray}
The dispersion relation (\ref{MIA-dispersion2}) indicates that the phase speed of the MIA waves decreases with the rise of the value of $\mu$ due to the reduction of the space charge electric field  by the presence of the PCD species.

Recently, due to this new linear feature of the MIA waves, the  subsonic solitary and shock waves  exist in a  dusty plasma system containing Maxwellian electron, cold ion, and PCD species \cite{Mamun21,Mamun20}. However, as indicated by (\ref{MIA-dispersion2}) the reduction of the MIA wave phase speed due to the presence of the PCD species can also make the MIA phase speed comparable with the ion thermal speed  $V_{Ti}=(k_BT_i/m_i)^{1/2}$ (where $T_i$ is the ion fluid
temperature) so that the effect of the ion-thermal pressure cannot be neglected. On the other hand, the electron species in many space environments does not always follow the Maxwellian velocity distribution function. This means that the dispersion relation (\ref{MIA-dispersion2}), and the works \cite{Mamun21,Mamun20} are valid only for two limiting cases: (i) cold ion ($T_i=0$) and (ii) Maxwellian electron species.

\begin{figure}[htb]
\centering
\includegraphics[width=80mm]{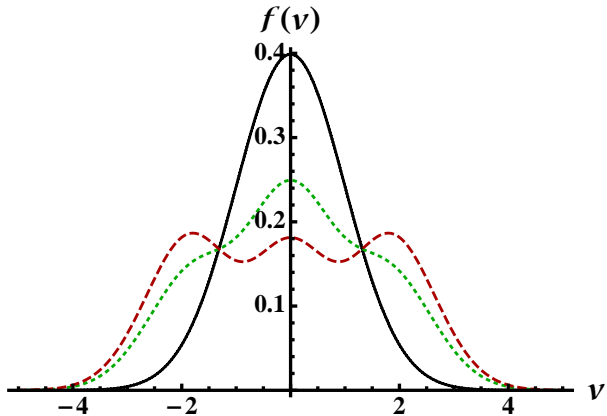}
\caption{How the different values of $\alpha$ [viz. $\alpha=0$ (solid curve),  $\alpha=0.2$ (dotted curve), and
$\alpha=0.4$ (dashed curve)] deviate from the curve corresponding to the Maxwellian distribution function to that corresponding to the Cairns distribution function \cite{Cairns95}. We note that $\phi=0.5$ is used.}
\label{F0}
\end{figure}

To overcome these two limitations, we consider (i) warm ion fluid ($T_i\ne 0$) \cite{Mamun97} and (ii) nonthermal electron species following the Cairns velocity distribution function in the form \cite{Cairns95}
\begin{equation}
f(v)= \frac{1+\alpha(v^2-2\phi)^2}{(1+3\alpha)\sqrt{2\pi}}\exp\left[-\frac{1}{2}(v^2-2\phi)\right],
\label{CDF}
\end{equation}
where $\phi$ is the MIA wave potential normalized by  $k_BT_e/e$,  $\alpha$ is a parameter determining the population of the fast (energetic) particles present in the plasma system under consideration. We note that the distribution function defined by (\ref{CDF}) is identical to the Maxwellian electron distribution function for
$\alpha=0$. How this nonthermal parameter $\alpha$ modifies the Maxwell distribution curve of the electron species is shown in figure \ref{F0}.
The effects of the Cairns nonthermal electron distribution ($\alpha$) and the ion fluid temperature ($\sigma=T_i/z_iT_e$) reduces the dispersion relation (\ref{MIA-dispersion2}) for the long wavelength MIA waves  to
\begin{eqnarray}
\frac{\omega}{kC_i}=\sqrt{\sigma+\frac{1+3\alpha}{(1+\mu)(1-\alpha)}}.
\label{MIA-dispersion3}
\end{eqnarray}
The dispersion relation
(\ref{MIA-dispersion3}) indicates that as $\alpha$ and $\sigma$ increase, the phase speed of the MIA waves increases and decreases the ion Landau damping \cite{Mamun11}.  This is due to the enhancement of the space charge electric field by the nonthermal electron species and by the thermal temperature of ion species.
The aim of this work is to investigate the combined effects of the PCD species, Cairns nonthermal electron species,  and  ion fluid temperature on the basic features of the MIA solitary (shock) waves in the electron-ion-PCD plasma system  by deriving the Korwege-de Vries (KDV) equaion \cite{Washimi66,Deeba12} (Burgers equation  \cite{Deeba12,Mamun09,Mamun19}).

The manuscript is structured in the following manner. The governing equations are provided in section \ref{GE}.
The basic features of solitary (shock) waves, are investigated by making use of the reductive perturbation method \cite{Washimi66,Deeba12,Mamun09,Mamun19}  in section \ref{KDVSW} (\ref{BSW}).  A brief discussion on our theoretical work is presented in section \ref{DIS}.

\section{Governing equations}
\label{GE}
We consider a plasma system containing nonthermally distributed  inertialess electron species [following  the Cairns distribution function represented by (\ref{CDF})], isothermal warm inertial ion species, and stationary PCD species. The nonlinear dynamics of the MIA waves  in this electron-ion-PCD plasma system is governed by  the normalized  equations in the form
\begin{eqnarray}
&&\frac{\partial n_i}{\partial t}+\frac{\partial}{\partial x} (n_iu_i) = 0,
\label{MIA-b1}\\
&&\frac{\partial u_i}{\partial t} + u_i \frac{\partial u_i}{\partial x} =-\frac{\partial \phi} {\partial x}-\frac{\sigma}{n_i} \frac{\partial n_i} {\partial x}+\eta\frac{\partial^2 u_i} {\partial x^2},
\label{MIA-b2}\\
&&\frac{\partial^2\phi}{\partial x^2}  =(1+\mu)n_e-n_i-\mu,
\label{MIA-b3}
\end{eqnarray}
where  $n_i$ ($n_e$) is the ion (electron) number density normalized by $n_{i0}$
($n_{e0}$);  $u_i$ is the ion fluid speed normalized
by $C_i$;  $x$ ($t$) is the space (time) co-ordinate normalized by $\lambda_D$ ($\omega_{pi}^{-1}$);
$\eta$ is the ion fluid  kinematic viscosity coefficient \cite{Mamun20} normalized by
$\omega_{pi}\lambda_D^2$.

The expression for $n_e$ is obtained by integrating
the non-thermal Cairns distribution function  [defined by (\ref{CDF})] over the whole velocity space, i.e.
\begin{eqnarray}
&&n_e=\int_{-\infty}^\infty f(v)dv=(1-\beta\phi+\beta\phi^{2})\exp(\phi),
\label{ne}
\end{eqnarray}
where $\beta=4\alpha/(1+3\alpha)$.  It  indicates  for  $\alpha=0$ that $n_e=\exp(\phi)$  yielding the Maxwellian  electron species.

\section{KDV Soltary Waves}
\label{KDVSW}
\begin{figure}[htb]
\centering
\includegraphics[width=80mm]{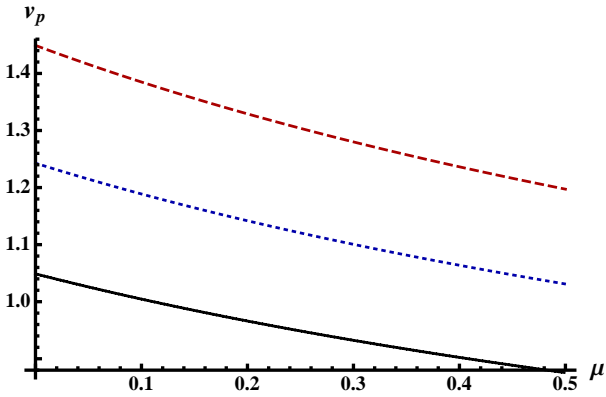}
\caption{The variation of $V_p$ with $\mu$  for different values of $\alpha$ [viz.
$\alpha=0$ (solid curve),  $\alpha=0.1$ (dotted curve), and  $\alpha=0.2$ (dashed curve)] and $\sigma=0.1$.}
\label{F1}
\end{figure}

\begin{figure}[htb]
\centering
\includegraphics[width=80mm]{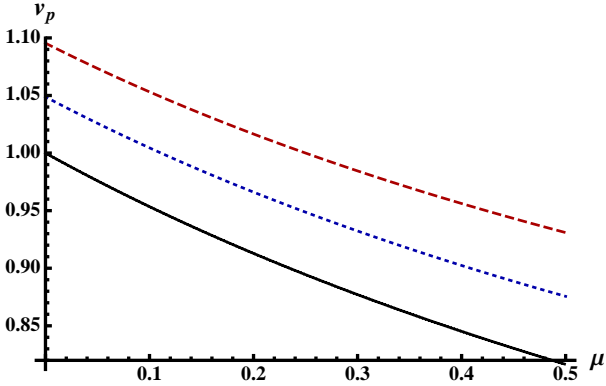}
\caption{The variation of $V_p$ with $\mu$ for different values of  $\sigma$  [viz.  $\sigma=0$ (solid curve),
$\sigma=0.1$ (dotted curve), and $\sigma=0.2$ (dashed curve) and $\alpha=0$.}
 \label{F2}
\end{figure}

To study the nature of the KDV solitary waves  in the electron-ion-PCD plasma system under consideration, we first  stretch the independent variables $x$ and $t$ as \cite{Washimi66,Deeba12}
\begin{eqnarray}
&&\zeta=\epsilon^{\frac{1}{2}}(x-{\cal V}_pt),
\label{str1}\\
&&\tau=\epsilon^{\frac{3}{2}}t,
\label{str2}
\end{eqnarray}
where $\epsilon$ is the small dimensionless expansion parameter, ${\cal V}_p$ is the MIA wave phase speed normalized by $C_i$ indicating ${\cal V}_p=\omega/kC_i$, and $\zeta$ ($\tau$) is normalized as $x$ ($t$) is.  We then expand the dependent variables $n_i$, $u_i$, and $\phi$ as \cite{Washimi66,Deeba12}
\begin{eqnarray}
&&n_i=1+\epsilon n_i^{(1)}+\epsilon^2 n_i^{(2)}+\cdot \cdot \cdot,
\label{expand1}\\
&&u_i=0+\epsilon u_i^{(1)}+\epsilon^2 u_i^{(2)}+\cdot \cdot \cdot,
\label{expand2}\\
&&\phi=0+\epsilon \phi^{(1)}+\epsilon^2 \phi^{(2)}+\cdot \cdot \cdot.
\label{expand3}
\end{eqnarray}
We now develop the equations in various powers of $\epsilon$.  To the lowest-order approximation,  we obtain
\begin{eqnarray}
&&n_i^{(1)}=\frac{1}{{\cal V}_p^{2}-\sigma}\phi^{(1)},
\label{L1}\\
&&u_i^{(1)}=\frac{{\cal V}_p}{{\cal V}_p^{2}-\sigma}\phi^{(1)},
 \label{L2}\\
&&{\cal V}_p=\sqrt{\sigma+\frac{1+3\alpha}{(1+\mu)(1-\alpha)}}.
 \label{L3}
\end{eqnarray}
This is the linear dispersion for the MIA waves in the  electron-ion-PCD plasma system, and
is identical to (\ref{MIA-dispersion3}) since ${\cal V}_p=\omega/kC_i$.
The interpretation of (\ref{MIA-dispersion3}) is verified graphically by displaying the figures \ref{F1} and \ref{F2} which show how ${\cal V}_p$ varies with $\mu$, $\alpha$, and $\sigma$.

To the next higher order approximation, we obtain
\begin{eqnarray}
&&\hspace*{-10mm}\frac{\partial n_i^{(1)}}{\partial \tau}
-{\cal V}_p\frac{\partial n_i^{(2)}}{\partial\zeta}+\frac{\partial u_i^{(2)}}{\partial\zeta}+
\frac{\partial}{\partial\zeta}\left[n_i^{(1)}u_i^{(1)}\right]=0,
\label{NL1}\\
&&\hspace*{-10mm}\frac{\partial u_i^{(1)}}{\partial \tau}-{\cal V}_p\frac{\partial u_i^{(2)}}{\partial \zeta}+\gamma u_i^{(1)}\frac{\partial u_i^{(1)}}{\partial \zeta}=-\frac{\partial \phi^{(2)}}{\partial \zeta}-\sigma\frac{\partial n_i^{(2)}}{\partial \zeta},
\label{NL2}\\
&&\hspace*{-10mm}\frac{\partial^2 \phi^{(1)}}{\partial \zeta^2}=(1+\mu)(1-\beta)\phi^{(2)}+\frac{1}{2}(1+\mu)\left[{\phi^{(1)}}\right]^{2}-n_i^{(2)},~~
\label{NL3}
\end{eqnarray}
where $\gamma=1+\sigma/{\cal V}_p^2$, which is obtained from (\ref{L1}) and (\ref{L2}).
We finally use (\ref{L1})$-$(\ref{NL3}) to obtain the KDV equation in the form
\begin{eqnarray}
\frac{\partial\phi^{(1)}}{\partial
\tau}+ {\cal A} \phi^{(1)} \frac{\partial
\phi^{(1)}}{\partial \zeta} + {\cal B} \frac{\partial^3 \phi^{(1)}}{\partial
\zeta^3} = 0,
\label{KDV}
\end{eqnarray}
where ${\cal A}$ and ${\cal B}$ denote the nonlinear and dispersion coefficients, which are, respectively, given by
\begin{eqnarray}
&&{\cal A}=\left[\frac{(3{\cal V}_p^{2}+\sigma)-(1+\mu)({\cal V}_p^{2}-\sigma)^{3}}{2{\cal V}_p({\cal V}_p^{2}-\sigma)}\right],
\label{A}\\
&&{\cal B}=\frac{({\cal V}_p^{2}-\sigma)^{2}}{2{\cal V}_p}.
\label{B}
\end{eqnarray}
\begin{figure}[htb]
\includegraphics[width=80mm]{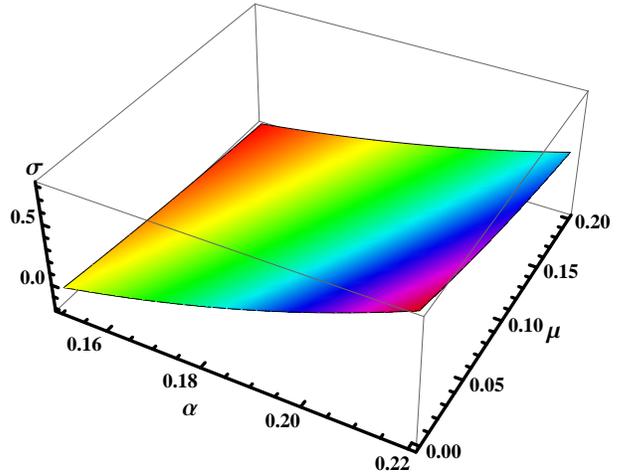}
\caption{${\cal A}=0$ surface plot showing the parametric regimes for the existence of the KDV solitary waves with  $\phi>0$ (parametric space \{$\mu$, $\alpha$, $\sigma$\}  above ${\cal A}=0$) surface) and $\phi<0$ (parametric space
\{$\mu$, $\alpha$, $\sigma$\}  below ${\cal A}=0$ surface).}
 \label{F3}
\end{figure}

\begin{figure}[htb]
\centering
\includegraphics[width=80mm]{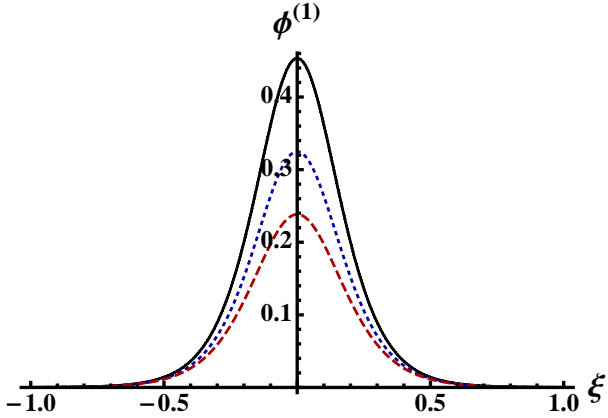}
\caption{The solitary potential profiles for ${\cal U}_0=0.1$, $\sigma=0.1$, $\alpha=0.1$, $\mu=0.2$ (solid curve), $\mu=0.3$ (dotted curve), and $\mu=0.4$ (dashed curve).}
\label{F4-so}
\end{figure}

\begin{figure}[htb]
\centering
\includegraphics[width=80mm]{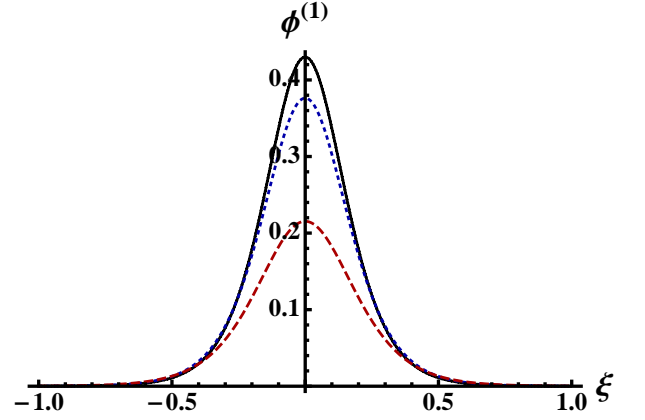}
\caption{The solitary potential profiles for ${\cal U}_0=0.1$, $\mu=0.3$, $\alpha=0.1$, $\sigma=0$ (solid curve),
$\sigma=0.2$ (dotted curve), and $\sigma=0.4$ (dashed curve).}
\label{F5-so}
\end{figure}

\begin{figure}[htb]
\centering
\includegraphics[width=80mm]{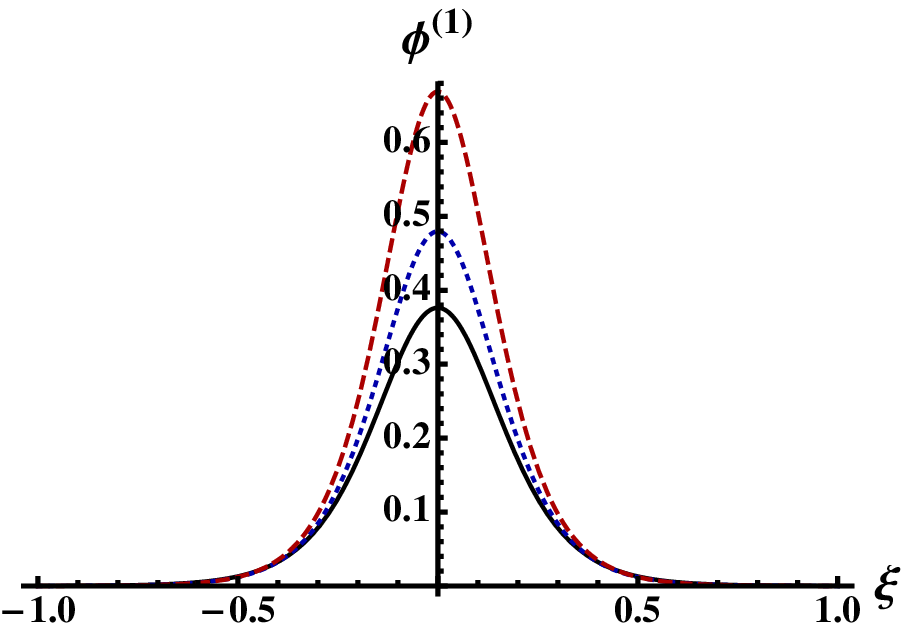}
\caption{The solitary potential profiles for ${\cal U}_0=0.1$,  $\mu=0.3$, $\sigma=0.1$, $\alpha=0.1$ (solid curve),
$\alpha=0.125$ (dotted curve), and $\alpha=0.15$ (dashed curve).}
\label{F6a-so}
\end{figure}

\begin{figure}[htp]
\centering
\includegraphics[width=80mm]{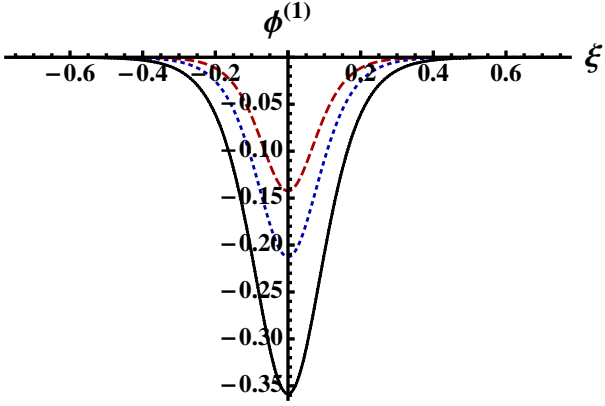}
\caption{The solitary potential profiles for ${\cal U}_0=0.1$,  $\mu=0.3$, $\sigma=0.1$, $\alpha=0.3$ (solid curve),
$\alpha=0.35$ (dotted curve), and $\alpha=0.4$ (dashed curve).}
\label{F6b-so}
\end{figure}

The stationary solitary wave solution of the KDV equation (\ref{KDV}) is derived by considering a moving frame
$\xi=\zeta-{\cal U}_0\tau$ (moving with a constant speed
${\cal U}_0$), where $\xi$ is normalized as $\zeta$ is, and ${\cal U}_0$  is normalized by $C_i$. Now, imposing the appropriate boundary conditions, viz. $\phi^{(1)}\rightarrow 0$,
$d\phi^{(1)}/d\zeta \rightarrow 0$, and $d^2\phi^{(1)}/d\zeta^2\rightarrow 0$ at $\xi\rightarrow \pm \infty$. These assumptions lead to the stationary solitary wave solution of the KDV equation (\ref{KDV}) in the form
\begin{eqnarray}
&&\phi^{(1)}=\phi_m^{(1)}{\rm
sech^2}\left(\frac{\xi}{\Delta}\right).
\label{KDV-sol}
\end{eqnarray}
where $\phi^{(1)}$ ($\Delta$) is the amplitude (width) of the MIA solitary waves, and are written as
\begin{eqnarray}
\phi_m^{(1)}=\frac{3{\cal U}_0}{{\cal A}}~~{\rm and}~~ \Delta=\sqrt{\frac{4{\cal B}}{{\cal U}_0}},
\label{Amp-Wid}
\end{eqnarray}
 respectively. It is clear from (\ref{KDV-sol})$-$(\ref{Amp-Wid}) that the subsonic and supersonic solitary waves (of amplitude $3{\cal U}_0/{\cal A}$ and width $\sqrt{4{\cal B}/{\cal U}_0}$) exist with $\phi^{(1)}>0$ [$\phi^{(1)}<0$] for ${\cal A}>0$ [${\cal A}<0$] since ${\cal U}_0>0$ and ${\cal B}>0$ are always valid, and that amplitude (width) of the MIA solitary waves increases (decreases) directly with ${\cal U}_0$
($\sqrt{{\cal U}_0}$).  The parametric regimes corresponding to ${\cal A}>0$ and ${\cal A}<0$ are shown by displaying the ${\cal A}=0$ surface plot in figure \ref{F3}.  To observe how the basic features (polarity, amplitude, and width) of the MIA solitary waves are modified by the number density of the stationary PCD species (represented by $\mu$), the ion fluid temperature (represented by $\sigma$), and the nonthermal parameter $\alpha$ (representing the population of the fast or energetic  electrons), we have graphically represented (\ref{KDV-sol}) for the different values of $\mu$,  $\sigma$, and $\alpha$.  The results are displayed in figures  \ref{F4-so}$-$\ref{F6b-so}.
\section{Burgers Shock Waves}
\label{BSW}
To study Burgers shock waves in the dissipative electron-ion-PCD plasma system under consideration,
we first stretch the independent variables $x$ and $t$ as \cite{Deeba12,Mamun09,Mamun19}
\begin{eqnarray}
&&\zeta=\epsilon(x-{\cal V}_pt),
\label{str3}\\
&&\tau=\epsilon^{2}t.
\label{str4}
\end{eqnarray}
We  then expand the dependent variables $n_i$, $u_i$, and $\phi$ by (\ref{expand1}), (\ref{expand2}), and (\ref{expand3}), respectively. We again develop the equations in various powers of $\epsilon$. To the lowest-order approximation, we obtain a linear set of equations, which are given by (\ref{L1})$-$(\ref{L3}).

To the next higher order approximation, we obtain
\begin{eqnarray}
&&\hspace*{-10mm}\frac{\partial n_i^{(1)}}{\partial \tau}
-{\cal V}_p\frac{\partial n_i^{(2)}}{\partial\zeta}+\frac{\partial u_i^{(2)}}{\partial\zeta}+
\frac{\partial}{\partial\zeta}\left[n_i^{(1)}u_i^{(1)}\right]=0,
\label{NL4}\\
&&\hspace*{-10mm}\frac{\partial u_i^{(1)}}{\partial \tau}-{\cal V}_p\frac{\partial u_i^{(2)}}{\partial \zeta}+u_i^{(1)}\frac{\partial u_i^{(1)}}{\partial \zeta}=-\frac{\partial \phi^{(2)}}{\partial \zeta}-\sigma\frac{\partial n_i^{(2)}}{\partial \zeta}\nonumber\\
&&~~~~~~~~~~~~~~~~~~~~~~~~~-\sigma n_i^{(1)}\frac{\partial n_i^{(1)}}{\partial \zeta}+\eta\frac{\partial^2 u_i^{(1)}} {\partial \zeta^2},
\label{NL5}\\
&&\hspace*{-10mm} 0=(1+\mu)(1-\beta)\phi^{(2)}+\frac{1}{2}(1+\mu)\left[{\phi^{(1)}}\right]^{2}-n_i^{(2)}.
\label{NL6}
\end{eqnarray}
We finally use (\ref{L1})$-$(\ref{L3}) and (\ref{NL4})$-$(\ref{NL6}) to obtain the Burgers equation in the form
\begin{eqnarray}
\frac{\partial\phi^{(1)}}{\partial
\tau}+ {\cal A} \phi^{(1)} \frac{\partial
\phi^{(1)}}{\partial \zeta} = {\cal C} \frac{\partial^2 \phi^{(1)}}{\partial
\zeta^2},
\label{Burgers}
\end{eqnarray}
where ${\cal A }$ (${\cal C}$) is the coefficient of nonlinearity (dissipation). The nonlinear coefficient ${\cal A}$ is by (\ref {A}), and the dissipative coefficient ${\cal C}$ is given by ${\cal C}=\eta/2$.
\begin{figure}[htb]
\centering
\includegraphics[width=80mm]{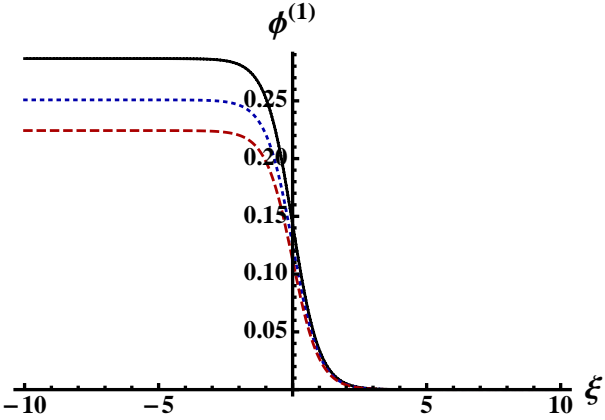}
\caption{The shock potential profiles for ${\cal U}_0=0.1$, $\sigma=0.1$, $\alpha=0.1$, $\eta=0.1$, $\mu=0.2$ (solid curve), $\mu=0.3$ (dotted curve), and $\mu=0.4$ (dashed curve).}
\label{F4-sh}
\end{figure}
\begin{figure}[htb]
\centering
\includegraphics[width=80mm]{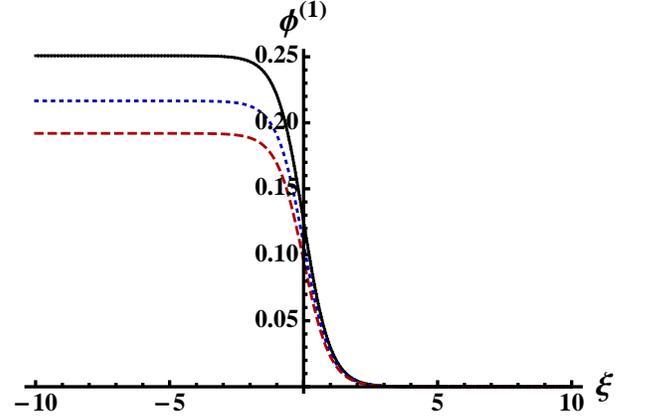}
\caption{The shock potential profiles for ${\cal U}_0=0.1$, $\mu=0.3$, $\alpha=0.1$, $\eta=0.1$, $\sigma=0.1$ (solid curve), $\sigma=0.2$ (dotted curve), and $\sigma=0.3$ (dashed curve).}
\label{F5-sh}
\end{figure}
\begin{figure}[htb]
\centering
\includegraphics[width=80mm]{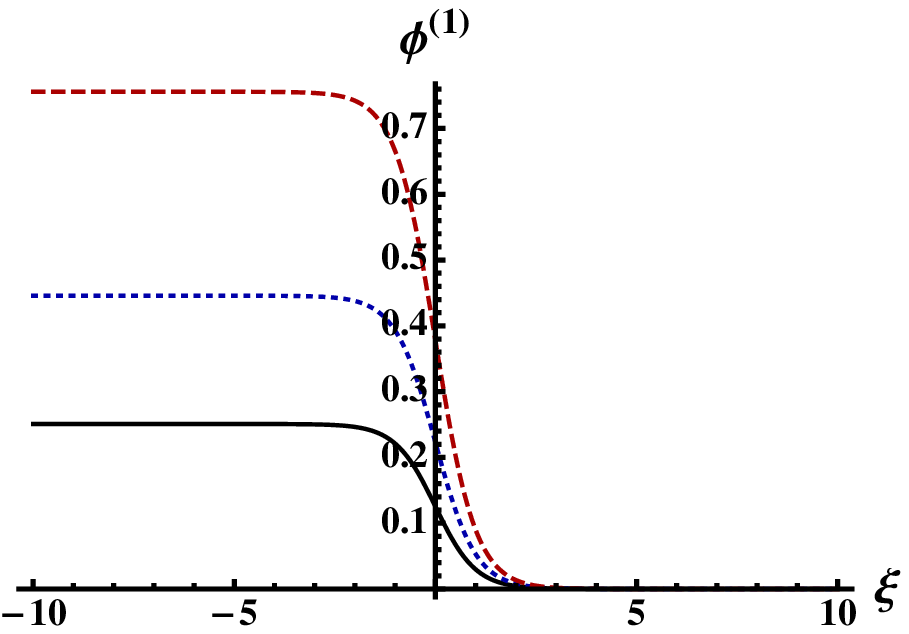}
\caption{The shock potential profiles for ${\cal U}_0=0.1$,  $\mu=0.3$, $\sigma=0.1$, $\eta=0.1$, $\alpha=0.1$ (solid curve), $\alpha=0.125$ (dotted curve), and $\alpha=0.15$ (dashed curve).}
\label{F6a-sh}
\end{figure}
\begin{figure}[htp]
\centering
\includegraphics[width=80mm]{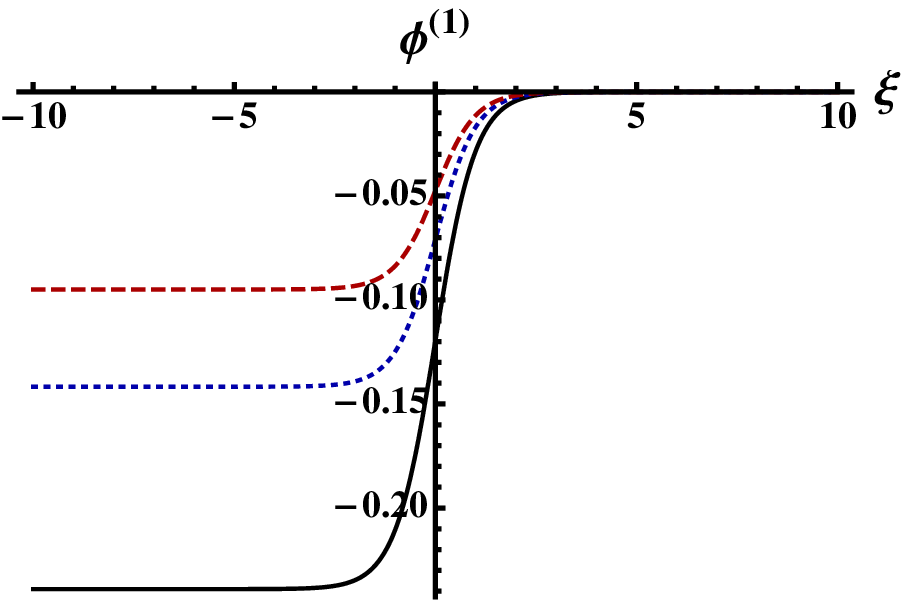}
\caption{The shock potential profiles for ${\cal U}_0=0.1$,  $\mu=0.3$, $\sigma=0.1$, $\eta=0.1$, $\alpha=0.3$ (solid curve), $\alpha=0.35$ (dotted curve), and $\alpha=0.4$ (dashed curve).}
\label{F6b-sh}
\end{figure}
To obtain the stationary shock wave solution of the Burgers equation (\ref{Burgers}), we again consider a moving frame moving with the speed ${\cal U}_0$, i.e. $\xi=\zeta-{\cal U}_0\tau$, and impose the appropriate boundary conditions ($\phi^{(1)}\rightarrow 0$ and $d\phi^{(1)}/d\xi\rightarrow 0$ at $\xi\rightarrow \infty$). These assumptions lead to the stationary shock wave solution of the Burgers equation (\ref{Burgers}) in the form
\begin{eqnarray}
&&\phi^{(1)}=\phi^{(1)}_m\Big[1-\tanh\bigg(\frac{\xi}{\Delta}\bigg)\Big],
\label{Burgers-sol}
\end{eqnarray}
where $\phi^{(1)}_m$ is the amplitude and $\Delta$ is the width of the shock waves, are given by
\begin{eqnarray}
&&\phi^{(1)}_m=\frac{{\cal U}_0}{\cal A},~~{\rm and}~~\Delta=\frac{\eta}{{\cal U}_0}.
\label{amp-thick}
\end{eqnarray}
It is obvious from (\ref{Burgers-sol}) and (\ref{amp-thick}) with figures \ref{F1}-\ref{F3}  that the subsonic and supersonic shock waves are formed with $\phi>0$  ($\phi<0$) for ${\cal A}>0$ (${\cal A}<0$). It also clear that the width of the shock waves is directly (inversely) proportional to $\eta$ (${\cal U}_0$).  However, to observe how the basic features (polarity, amplitude, and width) of the shock waves are modified by the number density of the stationary PCD species (represented by the parameter $\mu$), the ion fluid temperature (represented by the parameter $\sigma$), and the nonthermal parameter $\alpha$ (representing the population of the fast or energetic  electrons), we have graphically represented (\ref{Burgers-sol}) for the different values of $\mu$,  $\sigma$, and $\alpha$.  The results are displayed in figures \ref{F4-sh}$-$\ref{F6b-sh}.

\section{Discussion}
\label{DIS}
The dusty plasma system (containing inertialess nonthermal electron, warm inertial ion, and stationary PCD species) is considered. The roles of the stationary PCD species, ion fluid temperature, and nonthermal electrons in the linear and nonlinear propagation of the long (compared to $\lambda_D$) wavelength MIA  waves in such a plasma system are identified.  They can be pointed as follows:
\begin{itemize}
\item{The phase speed  of the long  wavelength MIA waves decreases with the rise of the number density and charge of the  stationary  PCD species. This is due  to the reduction of the space charge electric field (caused by the PCD species) in the electric field of the MIA waves in which inertia is provided by the positive ion fluid.}

\item{The phase speed of the long wavelength  MIA waves increases with the increase in the ion fluid temperature as well as in the fraction of the energetic electrons.  This is due  to the enhancement of the space charge electric field (caused by the more flexibility of the ion fluid and by the excess energy of
the energetic/fast particles) in the electric field of the MIA waves in which restoring force comes from the electron species.}

\item{The possibility for the formation of  the subsonic solitary and shock waves increases with the rise of the number density and charge of the stationary PCD species, since the latter reduces the phase speed  to its values of the subsonic range. But, this possibility decreases with the rise of the ion fluid temperature and with the rise of the fraction of the
fast/energetic electrons since the latter enhance the phase speed to the supersonic range.}

\item{The possibility for the formation of  the subsonic solitary and shock waves with negative potential  increases with the rise of  the fraction of the fast/energetic electrons. But this possibility decreases with the rise of the number density and charge of the stationary PCD species  as well as  with  the rise of the ion fluid temperature.}

\item{The amplitude of the solitary and shock waves decreases with rise of the values of number density and charge of the PCD species and with the ion fluid temperature, but increases with the increase in the faction of the fast/energetic electrons. The amplitude of the solitary waves is three times larger than that of the shock waves, but the nature of the variation of the shock and solitary wave amplitude are identical since the nonlinear coefficient in KDV and Burgers equations are identical.}

\item{The width of the solitary waves increases with rise of the values of the number density and charge of the PCD species and with the ion fluid temperature, but decreases with  the increase in the fraction of the fast/energetic electrons. }
\end{itemize}
It is important to mention that to derive KDV equation, we use the stretched or rescaled co-ordinates \cite{Washimi66}, which is valid only when the effect of dissipation is negligible in comparison with that of the dispersion, and that to derive Burgers equation, we use the stretched or rescaled co-ordinates \cite{Mamun09}, which is valid only when the effect of dispersion is negligible in comparison with that of the dissipation. We have used the reductive perturbation method with these KDV \cite{Washimi66} and Burgers \cite{Mamun09} stretching to derive the KDV and Burgers equations, respectively.

The limitation of the reductive perturbation method is that it is not valid for the arbitrary amplitude solitary and shock waves. To overcome this limitation, one has to develop a numerical code to simulate the basic equations (\ref{MIA-b1})$-$(\ref{MIA-b3}). This type of numerical simulation  will be able to show the time evolution of arbitrary amplitude solitary and shock waves.  This is, of  course, a challenging research problem, but beyond the scope of our present work. However, the results of our present work should be useful in understanding the ion-acoustic disturbances  in different space (viz. Earth's mesosphere \cite{Havnes96,Gelinas98,Mendis04},  upper part of ionosphere \cite{Dovner94,Cairns95}, cometary tails \cite{Horanyi96,Mamun04}, Jupiter's surroundings \cite{Tsintikidis96}, Jupiter's magnetosphere \cite{Horanyi93}, noctilucent clouds \cite{Markus99-NLCs}, etc.) and laboratory  \cite{Khrapak01,Fortov03,Davletov18} dusty plasma systems, where the PCD species coexists with the electron-ion plasmas.

Barkan {\it et al.} \cite{Barkan96} have performed a laboratory experiment on the propagation of the MIA waves in plasma system containing electron, ion, and negatively charged dust (NCD) species, and have shown that the presence of NCD species enhances the phase speed of the MIA waves. This experimental setup of Barkan {\it et al.} \cite{Barkan96} is expected to verify our result that the presence of the PCD species reduces the phase speed of the MIA waves by using PCD species instead of NCD species. On the other hand, Nakamura {\it et al.} \cite{Nakamura01,Nakamura99} have performed two laboratory experiments on the formation of solitary \cite{Nakamura01} and shock \cite{Nakamura99} waves in electron-ion-NCD plasmas. The basic features of the solitary and shock waves predicted in our present investigation can be tested by the experimental setup of Nakamura {\it et al.}  \cite{Nakamura01,Nakamura99} by replacing the NCD species by the PCD species.


\begin{thebibliography}{99}
\bibitem{Havnes96}
O. Havnes, J. Tr{\o}im, T. Blix, W. Mortensen, L. I. N{\ae}sheim,  E. Thrane, and T. T{\o}nnesen,  J.  Geophys. Res. {\bf 101}, 10839 (1996).

\bibitem{Gelinas98} L. J. Gelinas, K. A. Lynch, M. C. Kelley,
S. Collins, S. Baker, Q.  Zhou, and  J.  S. Friedman,
Geophys. Res. Lett. {\bf 25}, 4047 (1998).

\bibitem{Mendis04} D.  A.  Mendis, Wai-Ho Wong,  and M.  Rosenberg,
Phys. Scripta  {\bf T113}, 141 (2004).

\bibitem{Dovner94}  P.  O. Dovner,  A. I.  Eriksson and Bostr\"om,
 Geophys. Res. lett.  {\bf 21}, 1827 (1994).

 \bibitem{Cairns95} R. A. Cairns, A. A. Mamun, R. Bingham, R.  Bostr\"om, R. O. Dendy, C.
M. C. Nairn, and P. K. Shukla, Geophys. Res. Lett. \textbf{22}, 2709 (1995).

\bibitem{Horanyi96}  M.  Hor\'anyi,
Annu. Rev.  Astron.  Astrophys.  {\bf 34}, 383  (1996).

\bibitem{Mamun04} A.  A.  Mamun and P. K.  Shukla,
Geophys. Res.  Lett.  {\bf 31}, L06808 (2004).

\bibitem{Tsintikidis96} D. Tsintikidis,  D. A. Gurnett,  W.  S.  Kurth, and L. J. Granroth,
Geophys.  Res.  Lett. {\bf 23}, 997  (1996).

\bibitem{Horanyi93} M. Hor\'anyi, G. E. Morfill,  and E. Gr\"un,
Nature  London  {\bf 363}, 144 (1993).

\bibitem{Markus99-NLCs}  M.  Rapp and  F. J. L\"ubken,
Earth.  Planet.  Sp. {\bf 51}, 799 (1999).

\bibitem{Khrapak01}  S. A. Khrapak and G. Morfill,
Phys. Plasmas {\bf 8}, 2629 (2001).

\bibitem{Fortov03}  V.  E.  Fortov, A.  P.  Nefedov, O.  S.  Vaulina,
O.  F.  Petrov, I. E. Dranzhevski, A. M. Lipaev, and Y. P. Semenov,
New J. Physics  {\bf 5}, 102 (2003).

\bibitem{Davletov18} A.  E. Davletov, F.  Kurbanov, and Y. S. Mukhametkarimov,
Phys. Plasmas {\bf 25}, 120701 (2018).

\bibitem{Fortov98} V. E. Fortov, A. P.  Nefedov, O. S. Vaulina, A. M.  Lipaev, V.  I. Molotkov, A. A. Samaryan {\it et al.},
J.  Exp. Theor.  Phys.  {\bf 87}, 1087 (1998).

\bibitem{Chow93} V. W. Chow, D. A. Mendis, and M. Rosenberg,
J. Geophys. Res. {\bf 98},  19065  (1993).

\bibitem{Rosenberg95} M. Rosenberg and D. A. Mendis,
IEEE Trans. Plasma Sci. {\bf 23}, 117 (1995).

\bibitem{Rosenberg96} M. Rosenberg, D. A. Mendis, and D. P.  Sheehan,
IEEE Trans. Plasma Sci. {\bf 24},  1422  (1996).

\bibitem{Mamun21} A. A. Mamun,
Contrib. Plasma Phys. {\bf 61},  e202000192 (2021).

\bibitem{Mamun20} A. A.  Mamun and  B. E. Sharmin,
AIP Advances {\bf 10}, 125317 (2020).

\bibitem{Mamun97} A. A. Mamun, Phys. Rev. E {\bf 55} (2) (1997).

\bibitem{Mamun11} A. A. Mamun and P. K. Shukla,
J. Plasma Phys. {\bf 77},  437 (2021).

\bibitem{Washimi66} H. Washimi and T. Taniuti, Phys. Rev. Lett. \textbf{17}, 996 (1996).

\bibitem{Deeba12}  F.  Deeba, S. Tasnim, and A. A. Mamun,
IEEE Trans. Plasma Sci. {\bf 40},  2247 (2012).

\bibitem{Mamun09} A. A.  Mamun and R. A. Cairns,
Phys. Rev. E {\bf 79}, R055401 (2009).

\bibitem{Mamun19} A.  A.  Mamun,
Phys. Plasmas  {\bf 26}, 084501 (2019).

\bibitem{Barkan96} A. Barkan, N. D'Angelo, and R.  L. Merlino,
Planet. Space Sci. {\bf 44}, 239 (1996).

\bibitem{Nakamura01} Y.  Nakamura and A. Sharma,
Phys. Plasmas {\bf 8}, 3921 (2001).

\bibitem{Nakamura99} Y. Nakamura, H. Bailung, and P.  K. Shukla,
Phys.  Rev.  Lett. {\bf 83}, 1602 (1999).
\end{thebibliography}
\end{document}